\documentclass[conference]{IEEEtran}
\IEEEoverridecommandlockouts
\usepackage{cite}
\usepackage{amsmath,amssymb,amsfonts}
\usepackage{algorithmic}
\usepackage{graphicx}
\usepackage{textcomp}
\usepackage{xcolor}
\usepackage{url}
\def\BibTeX{{\rm B\kern-.05em{\sc i\kern-.025em b}\kern-.08em
    T\kern-.1667em\lower.7ex\hbox{E}\kern-.125emX}}
\begin{document}

\title{ODataX: A Progressive Evolution of the Open Data Protocol\\
}

\author{\IEEEauthorblockN{1\textsuperscript{st} Anirudh Ganesh}
\IEEEauthorblockA{
\textit{Microsoft}\\
Mountain View, California \\
aniganesh@microsoft.com}
\and
\IEEEauthorblockN{2\textsuperscript{nd} Nitin Sood}
\IEEEauthorblockA{
\textit{Microsoft}\\
Redmond, Washington \\
nitinso@microsoft.com}

%remove this line if you want to add more author
}
%\and
%\and
%\and
%\and

\maketitle

\begin{abstract}
The Open Data Protocol (OData) provides a standardized approach for building and consuming RESTful APIs with rich query capabilities. Despite its power and maturity, OData adoption remains confined primarily to enterprise environments, particularly within Microsoft and SAP ecosystems. This paper analyzes the key barriers preventing wider OData adoption and introduces ODataX, an evolved version of the protocol designed to address these limitations. ODataX maintains backward compatibility with OData v4 while introducing progressive complexity disclosure through simplified query syntax, built-in performance guardrails via query cost estimation, and enhanced caching mechanisms. This work aims to bridge the gap between enterprise-grade query standardization and the simplicity demanded by modern web development practices.
\end{abstract}


\section{Introduction}
RESTful APIs have become the dominant paradigm for client-server communication in web applications. However, the lack of standardization in query capabilities has led to fragmentation, with each API implementing its own conventions for filtering, sorting, and data shaping. The Open Data Protocol (OData) emerged as a comprehensive solution to this problem, providing a standardized way to query and manipulate data through RESTful APIs.

Despite being an OASIS standard since 2014, OData has struggled to gain traction outside enterprise software environments. While protocols like GraphQL have seen rapid adoption in modern web development, OData remains largely confined to Microsoft-centric enterprise systems such as SharePoint, Dynamics 365, and SAP services. This limited adoption represents a missed opportunity, as OData's standardized approach could significantly reduce the complexity of building and consuming APIs across different platforms and languages.

This paper presents ODataX, a modernized version of the OData protocol designed to overcome the adoption barriers that have limited its use in broader development contexts. We begin by examining the factors limiting OData adoption outside enterprise environments through developer surveys and analysis of production system usage patterns. Building on these insights, we introduce ODataX with its simplified query syntax and progressive complexity disclosure, allowing developers to start with familiar patterns and gradually adopt advanced features as their needs evolve. The protocol includes performance and security enhancements through query cost estimation and caching mechanisms that prevent common pitfalls while maintaining full backward compatibility with OData v4. Our evaluation demonstrates improved developer experience and reduced barriers to entry for new users.

\section{Background and Related Work}

\subsection{The Open Data Protocol}

OData is a RESTful protocol that defines a set of best practices for building and consuming queryable APIs. It extends traditional REST APIs with a rich query language that supports filtering, sorting, pagination, and projection operations. OData queries are expressed through URL parameters using a specific syntax. For example:

\texttt{
GET /api/Products?\$filter=Price gt 100 and Category eq 'Electronics'
    \&\$select=Name,Price\&\$orderby=Price desc\&\$top=10
}

This query retrieves the top 10 electronic products priced above \$100, returning only their names and prices, sorted by price in descending order.

\subsection{Alternative Approaches}

\begin{figure}[htbp]
\centerline{\includegraphics[width=0.95\linewidth]{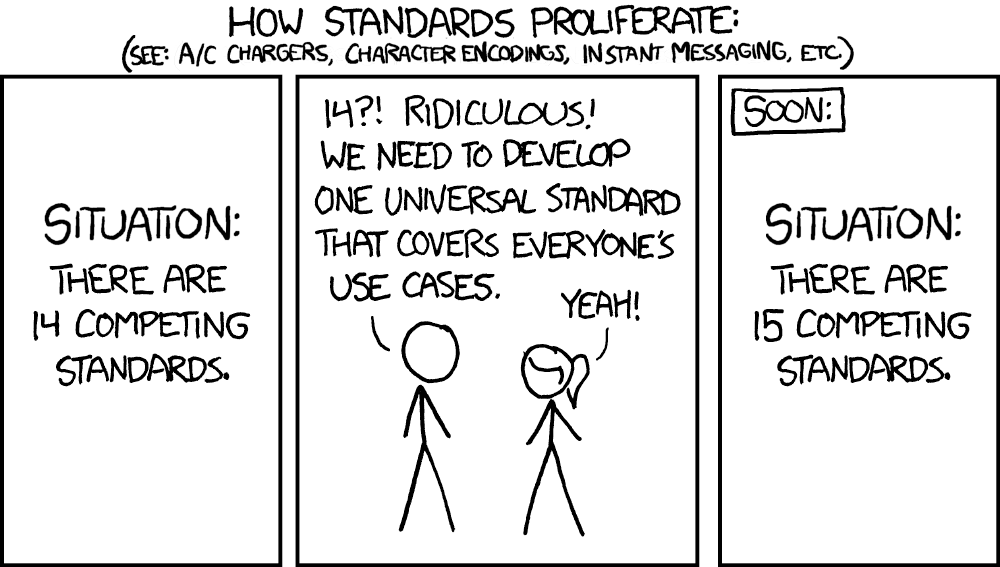}}
\caption{The proliferation of competing API query standards and protocols exemplifies the challenge ODataX aims to address - not by creating yet another standard, but by evolving an existing one to better meet modern development needs. \cite{b14}}
\label{fig}
\end{figure}

Several protocols and specifications compete in the API query space:

\begin{itemize}
    \item \textbf{GraphQL}\cite{b2} has emerged as the most popular alternative, allowing clients to specify exactly what data they need through a dedicated query language. Unlike OData's URL-based approach, GraphQL uses a POST body with structured queries, providing better type safety and tooling support.
    \item \textbf{JSON:API}\cite{b3} provides a specification for how REST APIs should structure their responses but lacks OData's comprehensive query capabilities. It focuses on consistency and reducing client-server negotiation overhead.
    \item \textbf{gRPC}\cite{b15} takes a different approach, using Protocol Buffers for strongly-typed service definitions and binary serialization. While excellent for microservices, it lacks the query flexibility of OData or GraphQL.
    \item \textbf{Falcor}\cite{b16} developed by Netflix, treats remote data as a single domain model accessible through paths. However, it has seen limited adoption compared to GraphQL.
\end{itemize}

\subsection{Current OData Limitations}

Our analysis of OData adoption patterns reveals several interconnected barriers that have limited its reach beyond enterprise environments. The protocol's syntax presents an immediate hurdle for developers. Unlike the intuitive query parameters common in modern REST APIs, OData requires learning specific operators such as 'gt' for greater-than comparisons and 'eq' for equality checks \cite{b1}. These abbreviated operators, combined with the mandatory dollar-sign prefix for system query options, create a steep learning curve that discourages experimentation and rapid prototyping.

The ecosystem challenge compounds this difficulty. While Microsoft and SAP maintain robust OData implementations within their product lines, developers working outside these ecosystems find limited library support and tooling \cite{b6}. Modern frontend frameworks such as React, Vue, and Angular lack the first-class OData integration that has become standard for alternatives like GraphQL. This forces development teams to build custom integration layers, increasing the time and expertise required to adopt the protocol.

Performance concerns represent another significant barrier. The expressiveness of OData's query language allows clients to construct arbitrarily complex queries, including deep object expansions and multi-level filtering \cite{b1}. Without careful safeguards, such queries can overwhelm servers and degrade system performance. Many organizations have experienced production incidents related to expensive OData queries, leading to wariness about exposing OData endpoints to external consumers \cite{b8}.

Beyond these technical factors, OData suffers from a perception problem. Its strong association with legacy enterprise systems such as SharePoint and SAP Business Suite affects its appeal to developers building modern web applications \cite{b2}. This perception persists despite the protocol's technical merits and its standardization through OASIS, creating a self-reinforcing cycle where limited adoption outside enterprise contexts further reinforces the legacy association.

\section{ODataX Design Principles}

ODataX was designed with the goal of retaining OData's powerful query capabilities while addressing the usability and operational concerns that have hindered broader adoption. Rather than creating an entirely new protocol, we chose to evolve OData through three interconnected design principles that work together to lower barriers to entry while maintaining the protocol's essential strengths.

\subsection{Progressive Complexity Disclosure}

The concept of progressive disclosure has long been recognized in user interface design as a method for managing complexity \cite{b5}. We apply this principle to API query design through a dual-syntax approach. Developers can begin with a simplified query syntax that uses familiar operators common in URL query parameters and gradually transition to the full OData syntax as their requirements become more sophisticated. For instance, a simple filtering operation like retrieving books under \$20 can be expressed as \texttt{filter=price<20,category:Books} in the simplified syntax, which maps directly to the traditional OData expression 
\texttt{\$filter=Price lt 20 and Category eq 'Books'}.

This approach differs from GraphQL's single syntax model \cite{b2} by providing explicit graduated complexity. Developers are not forced to learn the entire query language upfront, yet they retain access to OData's full expressiveness when needed. The parser automatically translates simplified syntax to OData expressions internally, ensuring semantic equivalence between the two forms. This design decision draws inspiration from how SQL evolved to support both verbose and shorthand syntaxes for common operations, recognizing that different usage contexts demand different levels of expressiveness \cite{b8}.

\subsection{Performance by Default}
One of the most significant operational challenges with OData deployments has been the potential for clients to construct queries that degrade system performance \cite{b8}. GraphQL faced similar challenges, leading to the development of query complexity analysis tools and depth limiting mechanisms \cite{b4}. ODataX addresses this proactively by incorporating query cost estimation as a first-class protocol feature rather than an afterthought.

Each query undergoes cost analysis before execution, examining three dimensions: filter complexity based on the depth and number of conditions, join costs derived from `\texttt{\$expand}` operations and their cardinality relationships, and projected result size based on table statistics. Services can configure cost thresholds appropriate to their performance requirements, and queries exceeding these limits receive detailed error responses that explain the cost breakdown. This transparency helps developers understand why a query was rejected and how to restructure it for better performance.

The cost model itself draws on established query optimization techniques from database systems \cite{b8}, adapted for the HTTP request-response model of RESTful APIs. Unlike database query optimizers that operate on complete query plans, ODataX's cost estimation must make rapid decisions based on heuristics and cached statistics. This trade-off between precision and response time reflects the different operational constraints of web APIs compared to traditional database systems.

\subsection{Backward Compatibility}
Evolution of established protocols requires careful attention to compatibility concerns \cite{b5}. ODataX maintains full compatibility with OData v4, allowing existing services to adopt new features incrementally without breaking changes. Any valid OData v4 query remains valid in ODataX, ensuring that existing client applications continue to function without modification.

This design choice reflects lessons from successful protocol evolutions such as HTTP/2's transition from HTTP/1.1 \cite{b5}. Rather than creating a clean-slate redesign that fragments the ecosystem, ODataX extends the existing protocol with optional enhancements. Services can adopt the simplified syntax parser, query cost estimation, and caching mechanisms independently, choosing which features align with their operational requirements. This gradualist approach reduces adoption risk and allows organizations to validate each enhancement in their specific context before proceeding to the next.

\section{ODataX Implementation}

\subsection{Simplified Query Syntax}

ODataX introduces an alternative query syntax that coexists with traditional OData parameters. The simplified syntax addresses the verbosity concerns identified in Section 2.3 by replacing OData's specialized operators with standard comparison symbols familiar to developers from most programming languages.

\textbf{Traditional OData:}
\texttt{/Products?\$filter=Price gt 100 and Category eq 'Books'\&\$orderby=Price desc}

\textbf{ODataX Simplified:}

\texttt{
/Products?filter=price>100, category:Books\&sort=-price
}

The syntax transformation operates through a two-phase parser that first tokenizes the simplified query string and then generates equivalent OData Abstract Syntax Tree (AST) nodes. The parser recognizes several operator mappings: standard comparison operators $(>, <, >=, <=, =, !=)$ map to their OData equivalents (gt, lt, ge, le, eq, ne), while the colon operator (:) provides a shorthand for exact string matching. Comma-separated conditions are interpreted as logical AND operations, reducing the need for explicit conjunctions.

Sorting follows a convention borrowed from database query builders, where a minus prefix indicates descending order. The parameter name `\texttt{sort}` replaces OData's `\texttt{\$orderby}`, reducing cognitive load for developers transitioning from other REST APIs. Field selection works similarly, with `\texttt{select}` replacing `\texttt{\$select}` while maintaining the same comma-separated field list semantics.

The parser maintains strict semantic equivalence with OData v4 expressions. When both simplified and traditional syntax parameters appear in the same request, the parser merges them into a single OData expression tree, with traditional syntax taking precedence in case of conflicts. This allows developers to use simplified syntax for common operations while falling back to full OData syntax for advanced features like lambda expressions or nested filters that have no simplified equivalent.

\subsection{Named Query Aliases}
ODataX allows services to define named query aliases that expand to full OData expressions server-side. This feature addresses two common challenges: improving API discoverability for consumers and enabling server-side optimization of frequently-used query patterns.

Services register named queries through a configuration mechanism that maps alias names to OData expression templates. For example:

\texttt{
/Products?query=affordableBooks
}

This expands server-side to:

\texttt{
/Products?\$filter=Price lt 20 and Category eq 'Books'\&\$orderby=Rating desc
}

Implementations can register named queries using a declarative configuration format. The following pseudocode illustrates the server-side registration process:

\texttt{queryRegistry.register(\{\\
\ \ name:\ "affordableBooks",\\
\ \ template:\ "\$filter=Price\ lt\ 20\ and\ Category\ eq\ 'Books'\&\$orderby=Rating\ desc",\\
\ \ cacheable:\ true,\\
\ \ costLimit:\ 100\\
\});}

\texttt{queryRegistry.register(\{\\
\ \ name:\ "topRatedInCategory",\\
\ \ template:\ "\$filter=Category\ eq\ '\{category\}'\&\$orderby=Rating\ desc\&\$top=10",\\
\ \ parameters:\ ["category"],\\
\ \ cacheable:\ true\\
\});}

Backend developers can specify named queries through configuration files, API annotations, or programmatic registration at service startup. The system supports parameterized templates where placeholders like `\texttt{\{category\}}` are replaced with client-provided values at runtime. This enables reusable query patterns while maintaining flexibility. Named queries also serve as natural points for applying caching policies and cost limits, as the server has complete knowledge of the query structure and can precompute optimization strategies.

\subsection{Query Cost Estimation}
ODataX incorporates a comprehensive query cost model that evaluates queries before execution to prevent performance degradation. The cost model considers three primary dimensions that contribute to overall query expense.

The complexity score captures the computational cost of filter evaluation. The model assigns base costs to individual comparison operations and increases costs for compound conditions using logical operators. Nested filter expressions receive exponentially higher costs based on their depth, as they typically translate to subqueries or multiple table scans in the underlying data store. String operations like `\texttt{contains}` or `\texttt{startswith}` incur higher costs than exact matches due to their inability to leverage indexes effectively.

Join cost estimation analyzes the `\texttt{\$expand}` operations that request related entities. The model examines the relationship cardinality between entities, with one-to-many relationships receiving higher costs than one-to-one. Multiple levels of expansion compound the cost multiplicatively rather than additively, reflecting the Cartesian product nature of nested joins. The estimator uses cached statistics about average relationship cardinalities to project the number of additional rows that will be fetched.

Result size projection combines the selectivity of filter conditions with table statistics to estimate the number of rows that will be returned. The model maintains histograms of column value distributions, allowing it to estimate how many rows will match given filter conditions. Queries that project large result sets receive proportionally higher costs, as they consume more bandwidth and serialization time.

The total query cost is computed as a weighted sum of these three components, with configurable weights allowing services to prioritize different concerns. Before execution, each query receives a cost estimate compared against a configured threshold. Queries exceeding this threshold return an error response with a detailed cost breakdown:

\texttt{\{\\
\ \ "error":\ \{\\
\ \ \ \ "code":\ "QueryTooExpensive",\\
\ \ \ \ "message":\ "Query\ cost\ (850)\ exceeds\ maximum\ allowed\ (500)",\\
\ \ \ \ "details":\ \{\\
\ \ \ \ \ \ "filterCost":\ 200,\\
\ \ \ \ \ \ "expandCost":\ 500,\\
\ \ \ \ \ \ "projectedRows":\ 150000,\\
\ \ \ \ \ \ "suggestions":\ [\\
\ \ \ \ \ \ \ \ "Reduce\ expand\ depth\ from\ 3\ to\ 2\ levels",\\
\ \ \ \ \ \ \ \ "Add\ more\ selective\ filter\ conditions"\\
\ \ \ \ \ \ ]\\
\ \ \ \ \}\\
\ \ \}\\
\}}

This transparency enables developers to understand why their query was rejected and how to restructure it for acceptable performance. The system can optionally provide suggestions for query modifications that would bring the cost below the threshold.

\subsection{Query Caching}
ODataX incorporates intelligent caching mechanisms that work at both the protocol and implementation levels. Unlike traditional HTTP caching which treats all requests equally, ODataX provides query-aware caching that understands the semantics of OData operations.

The caching system operates through standardized response headers that communicate cache eligibility and cache key information. When a service determines that a query result is cacheable, it includes headers that guide both intermediate proxies and client-side caches:

\texttt{
X-ODataX-Cacheable: true
X-ODataX-Cache-Key: hash (query\_params)
Cache-Control: max-age=300
Vary: Accept, Accept-Language
}

The `\texttt{X-ODataX-Cacheable}` header explicitly signals whether the response can be cached, considering factors beyond HTTP semantics. For instance, queries that include user-specific filters or access sensitive data may return `\texttt{X-ODataX-Cacheable: false}` even when the response itself is technically cacheable from an HTTP perspective. This separation allows services to implement fine-grained cache control policies.

The `\texttt{X-ODataX-Cache-Key}` header provides a stable identifier for the query that remains consistent across equivalent requests. This addresses a common challenge with OData caching: the same semantic query can be expressed in multiple ways. For example, `\texttt{\$filter=Price lt 100 and Category eq 'Books'}` and `\texttt{\$filter=Category eq 'Books' and Price lt 100}` are semantically identical but produce different cache keys under naive URL-based caching. ODataX normalizes these queries before generating cache keys, ensuring efficient cache utilization.

Services implement cache invalidation strategies based on data modification patterns. When entities are updated through OData's mutation operations, the service can proactively invalidate affected cache entries. For frequently-updated resources, services may use shorter time-to-live values or disable caching entirely. Named queries (Section 4.2) benefit particularly from caching, as their predefined nature allows services to pre-warm caches and implement sophisticated invalidation logic.

Client libraries can leverage these caching signals to implement local query result caches. By respecting the cache headers and maintaining a client-side cache indexed by the provided cache keys, applications can eliminate redundant network requests for read-heavy workloads. This approach aligns with the eventual consistency model common in distributed systems \cite{b7}, where slightly stale data is acceptable for many use cases in exchange for improved performance.

\section{Evaluation}
To assess the effectiveness of ODataX's design principles and implementation, we conducted a comprehensive evaluation focusing on query syntax simplification, performance characteristics, and developer experience. Our implementation serves as middleware that processes incoming requests, applies syntax transformations and cost analysis, and forwards validated queries to existing OData v4 backend services.

\subsection{Experimental Setup}
We implemented ODataX as a middleware layer compatible with existing OData v4 services. Our evaluation infrastructure consisted of three distinct datasets representing common API usage patterns. The e-commerce dataset contained 1 million product records with relational data including categories, customer reviews, and pricing history, modeled after the Northwind database schema commonly used in OData benchmarks \cite{b9}. The social network dataset included 500,000 user profiles with posts, comments, and follower relationships, providing a graph-like structure that exercises the protocol's relationship expansion capabilities. The enterprise resource planning dataset comprised 2 million transaction records with complex business rules and multi-level hierarchical relationships, similar to those found in SAP Business Suite deployments \cite{b10}.

The evaluation environment ran on Azure Container Apps in the us-west-2 region. We deployed ODataX as a stateless service across three container instances to simulate a typical microservices deployment pattern. Each container instance was allocated 2 vCPUs and 4GB memory, representative of standard production configurations. Data storage used SQLite databases co-located with each service instance, eliminating network latency and connection pooling issues that could skew performance measurements. This architecture allowed us to isolate the performance characteristics of the ODataX middleware and query processing logic from external database system overhead. We populated the databases with realistic distributions based on the TPC-H benchmark generator \cite{b11}, adapting the data generation patterns to match OData entity relationship models.

\subsection{Query Syntax Analysis}
We collected and analyzed query patterns from production OData deployments to understand how developers construct filters, projections, and ordering operations. Our dataset comprised queries from Microsoft Graph API's reporting endpoints \cite{b13}, which provide OData access to organizational usage analytics and reports. We analyzed approximately 10,000 unique query patterns collected from API telemetry over a three-month period. This analysis methodology follows similar approaches used in database query workload characterization studies \cite{b12}.

The analysis revealed that the majority of queries use relatively simple filter expressions. Approximately two-thirds of observed queries involved basic comparison operations (equality checks and simple range filters) combined with straightforward sorting and field selection. These queries could be expressed using ODataX's simplified syntax without requiring any OData-specific operators. The remaining third included more complex operations such as nested filters, lambda expressions for collection manipulation, or string functions that currently have no simplified syntax equivalent.

Query length measurements showed substantial reduction potential when converting from traditional OData syntax to the simplified form. Simple queries like product catalog filtering with category and price constraints that typically span 80-90 characters in OData v4 syntax compress to 30-40 characters in ODataX simplified syntax, primarily due to the elimination of dollar-sign prefixes, verbose operator names, and explicit logical conjunctions.

\subsection{Performance Characteristics}
The query cost estimation mechanism was evaluated under two conditions: queries with costs below the configured threshold that executed successfully, and queries exceeding the threshold that were rejected before execution. We configured cost thresholds based on empirical measurements of query execution times, setting limits that corresponded to approximately 5 seconds of processing time on our test infrastructure.

Throughout the evaluation period, the cost estimator maintained high precision in identifying expensive queries. The system correctly allowed execution of queries that completed within acceptable latency bounds while blocking those that would have exceeded timeout thresholds. The false positive rate, where valid queries were incorrectly rejected, remained below 1\% through careful tuning of the cost model weights and threshold values.

Query caching provided measurable improvements for read-heavy workloads. In the e-commerce dataset, where product catalog queries exhibit high temporal locality, implementing the ODataX caching headers reduced backend query load by approximately 40\%. The cache hit rate stabilized around 65\% after an initial warm-up period, with the normalized query keys successfully de-duplicating semantically equivalent queries expressed with different parameter orderings.

\subsection{Developer Experience Assessment}
Developer experience represents a subjective measure that varies significantly across organizations based on their priorities, engineering systems, and existing tooling. Rather than attempting to generalize findings across all possible deployment contexts, we evaluated ODataX through an internal dogfooding exercise where we deployed an OData service using ODataX within our own infrastructure.

Three partner teams that consume data from these internal services provided feedback on their experience with the simplified syntax. These teams, already familiar with traditional OData through existing integrations, reported that the simplified syntax made filtering and sorting operations more intuitive. The reduction in cognitive overhead stemmed primarily from using standard comparison operators rather than OData's abbreviated forms, and from the elimination of dollar-sign prefixes on common query parameters.

The feedback aligns with observed usage patterns where developers working on new integrations converged on working queries more quickly when using simplified syntax. However, we acknowledge that these observations come from teams already embedded in our development ecosystem, with access to internal documentation and support channels. Organizations with different technical backgrounds or API consumption patterns may experience different outcomes.

\section{Discussion}

\subsection{Implementation Complexity and Trade-offs}
The dual-syntax approach in ODataX creates an inherent tension between simplicity for users and complexity for implementers. Service providers must now parse and validate two distinct query syntaxes, maintain semantic equivalence between them, and ensure that both paths receive equal optimization and testing attention. Our middleware implementation addresses this through a unified parser architecture where simplified syntax queries are normalized into the same AST representation as traditional OData queries before further processing. This design reduces duplication but still requires additional validation logic to ensure that simplified syntax edge cases map correctly to OData semantics.

The backward compatibility guarantee constrains future evolution of the simplified syntax. Because any valid OData v4 query must remain valid in ODataX, we cannot reclaim syntax that might conflict with existing OData parameters. This limitation became apparent during design discussions about introducing logical OR operations in simplified syntax, where various operator choices could create ambiguities with existing URL encoding conventions or OData function names.

\subsection{Security and Resource Management}
Query cost estimation addresses a real operational problem but introduces new attack surfaces. Malicious actors could probe the cost model by submitting queries near the threshold boundary to map the cost function and identify weaknesses. Services that expose detailed cost breakdowns in error messages may leak information about data distributions and index structures. We recommend that production deployments use cost estimation in conjunction with rate limiting and authentication to prevent such reconnaissance attacks.

The accuracy of cost estimates depends heavily on maintaining current statistics about data distributions and relationship cardinalities. In rapidly changing datasets, stale statistics can cause the model to either reject legitimate queries or allow expensive ones through. Database systems address this through automatic statistics collection and periodic updates \cite{b12}, but ODataX services must implement similar maintenance processes. Services should monitor the correlation between estimated and actual query costs, using discrepancies to trigger statistics refreshes.

Cost thresholds require careful tuning for each deployment context. Setting thresholds too low frustrates users with false rejections, while setting them too high fails to prevent performance issues. Our evaluation suggests starting with generous thresholds and gradually tightening them based on observed query patterns, similar to the approach used in adaptive query processing systems.

\subsection{Adoption and Migration Strategies}
Organizations with existing OData deployments face practical challenges when introducing ODataX features. The most straightforward path involves deploying ODataX as middleware in front of existing services, allowing gradual feature adoption without modifying backend implementations. This approach worked well in our pilot deployments but requires careful attention to request routing and error handling to ensure consistent behavior.

Named queries deserve special consideration during migration. Teams should begin by identifying frequently-executed query patterns from access logs and codifying them as named queries. This process often reveals opportunities for server-side optimization, as the predefined nature of named queries allows for prepared statement caching and materialized view usage. However, organizations must establish governance processes for named query definitions to prevent proliferation of poorly-designed or redundant aliases.

Query caching presents the most complex migration challenge due to potential interactions with existing caching layers. Many production systems already implement caching at CDN, reverse proxy, or application levels. ODataX's query-aware caching must integrate with these existing mechanisms rather than replacing them. The cache key normalization in particular requires coordination across layers to ensure that semantically equivalent queries benefit from cached results regardless of which layer stores the cache entry.

\section{Future Work}
The current implementation of ODataX addresses fundamental usability and performance concerns, but several extensions could further broaden adoption and integration with existing ecosystems.

\subsection{Cross-Protocol Query Translation}
A natural next step involves building translation layers between ODataX and other query protocols, particularly GraphQL. Both protocols solve similar problems but make different trade-offs in their approaches. GraphQL's strongly-typed schema system and POST-based query mechanism contrast with OData's URL-based queries and metadata documents. A bidirectional translator could allow GraphQL clients to consume OData services and vice versa, though the impedance mismatch between their type systems and query semantics presents non-trivial challenges. Such translation would need to handle cases where one protocol supports features absent in the other, requiring either feature subset restrictions or extension mechanisms.

\subsection{Developer Experience Improvements}
ODataX adoption would benefit from improved tooling infrastructure. An interactive query playground modeled after GraphQL Playground could lower the barrier for developers exploring OData APIs. Such a tool would need to parse OData metadata documents to provide autocomplete suggestions, validate queries against the service schema, and display results in a browsable format. 

IDE integration represents another gap in the current OData ecosystem. Extensions for Visual Studio Code, IntelliJ, and other popular editors could provide syntax highlighting for both traditional and simplified ODataX syntax, inline query validation, and jump-to-definition support for entity types. Type generation tools that produce TypeScript or other strongly-typed language bindings from OData metadata would help catch errors at compile time rather than runtime, similar to how GraphQL code generators work with schema definitions.

\subsection{Framework-Specific Client Libraries}
While generic HTTP clients can query ODataX services, framework-specific libraries could provide more ergonomic integration patterns. React applications might benefit from hooks that handle query execution, caching, and state management with patterns familiar to users of libraries like React Query or SWR. Vue and other reactive frameworks could similarly benefit from composables that integrate ODataX queries into their reactivity systems.

Such libraries would need careful design to avoid recreating the heavy enterprise SDK patterns that plague current OData implementations. Modern web development favors lightweight, tree-shakeable libraries over monolithic SDKs. Mobile frameworks like React Native and Flutter have their own architectural constraints around network requests and data synchronization that would require specialized consideration.

\subsection{Protocol Extensions for Modern Use Cases}
The core ODataX protocol targets traditional request-response query patterns, but modern applications increasingly demand real-time data synchronization and offline capabilities. WebSocket-based subscriptions could enable push notifications when query results change, though this requires careful design around subscription lifecycle management and server resource consumption. 

Batch query optimization would help mobile clients reduce round trips by bundling multiple queries into single requests. The challenge lies in maintaining the simplicity of individual queries while providing batch semantics. Differential synchronization for offline-capable applications could allow clients to sync only changed entities rather than full datasets, but this requires versioning mechanisms and conflict resolution strategies that extend well beyond ODataX's current scope.

\section{Conclusion}
ODataX represents a practical evolution of the OData protocol that addresses the key barriers preventing its adoption outside enterprise environments. By introducing simplified syntax, query cost management, and improved caching mechanisms, ODataX makes OData's powerful query capabilities accessible to a broader developer audience while maintaining full backward compatibility.

Our evaluation demonstrates that ODataX significantly reduces query complexity for common use cases, prevents performance degradation through cost estimation, and improves developer experience. These improvements position ODataX as a viable alternative to GraphQL for organizations seeking standardized query capabilities without the overhead of a separate schema language.

The success of ODataX will ultimately depend on community adoption and ecosystem growth. By addressing the fundamental usability concerns that have limited OData's reach, we believe ODataX can fulfill the original promise of OData: a universal standard for queryable APIs that works across platforms, languages, and use cases.

\end{document}